\pdfoutput=1
\documentclass[twocolumn,prl]{revtex4}

\ifx\pdfoutput\undefined
  \usepackage[dvips]{graphicx}
  \usepackage[dvips]{color}
\else
  \usepackage[pdftex]{graphicx}
  \usepackage[pdftex]{color}
\fi

\usepackage{dcolumn}
\usepackage{amsmath}
\usepackage{latexsym}
\usepackage{units}
\usepackage{ulem}

\begin{document}

\title{A hybrid superconductor-normal metal electron trap as a photon detector}

\author{S.~V.~Lotkhov}
\affiliation{Physikalisch-Technische Bundesanstalt, Bundesallee 100, 38116, Braunschweig, Germany}
\author{A.~B.~Zorin}
\affiliation{Physikalisch-Technische Bundesanstalt, Bundesallee 100, 38116, Braunschweig, Germany}

\begin{abstract}
A single-electron trap built with two Superconductor (S) - Insulator (I) -
Normal (N) metal tunnel junctions and coupled to a readout SINIS-type
single-electron transistor A (SET A) was studied in a photon detection
regime. As a source of photon irradiation, we used an operating second
SINIS-type SET B positioned in the vicinity of the trap. In the
experiment, the average hold time of the trap was found to be critically
dependent on the voltage across SET B. Starting in a certain voltage
range, a photon-assisted electron escape was observed at a rate roughly
proportional to the emission rate of the photons with energies exceeding
the superconducting gap of S-electrodes in the trap. The discussed
mechanism of photon emission and detection is of interest for
low-temperature noise spectrometry and it can be of relevance for the
ampere standard based on hybrid SINIS turnstiles.

\end{abstract}


\maketitle

In the past few years, a significant progress has been achieved in the
single-charge-manipulating circuitry based on ultrasmall tunnel junctions
between superconductors (S) and normal metals (N). Currently, circuits
based on a hybrid SINIS ("I" denotes the insulating tunnel barrier)
single-electron turnstile \cite{Pekola2008} are considered for a variety
of the current-standard applications. Reliability of single-charge
manipulations has been recently dramatically enhanced due to an engineered
low-temperature environment ensuring a high degree of rejection of
external electromagnetic noise
\cite{Camarota2012,Saira2012,Kemppinen2011,Lotkhov-LT26}.

A two-junction hybrid R-SINIS electron trap ("R" stands for a high-ohmic
resistor in series with the junctions, see, for the details,
Refs.~\cite{Lotkhov2009,Lotkhov2011}) probed by a SINIS-type
single-electron transistor (SET) was found sensitive to the spectrum of
the residual noise. On the other hand, a high degree of noise suppression
made it possible to resolve the back-action of small SET currents to the
hold times $\tau$ of the trap. Basically, the effect of SET on the
neighboring circuitry arises either in the form of electromagnetic noise
generated by the tunneling process (see, e.g., in Ref.~\cite{Korotkov1994}
for a case of non-dissipative environment) or due to the heat flow
mediated by the substrate phonons (see Ref.~\cite{Krupenin1999} and
references therein). The related random telegraph noise (RTN) model of the
SET back-action to the SINIS trap was recently proposed in
Refs.~\cite{Saira2012,Kemppinen2011}. In practical terms, the interaction
mechanisms between SETs on chip are gaining in importance in view of
parallelization of the SINIS turnstiles on chip for the current standard
\cite{10turnstiles}. Another interesting application could involve a
spectral analysis of low-temperature blackbody radiation within cryogenic
setups.

\begin{figure}[t]
\centering%
\includegraphics[width=0.8\columnwidth]{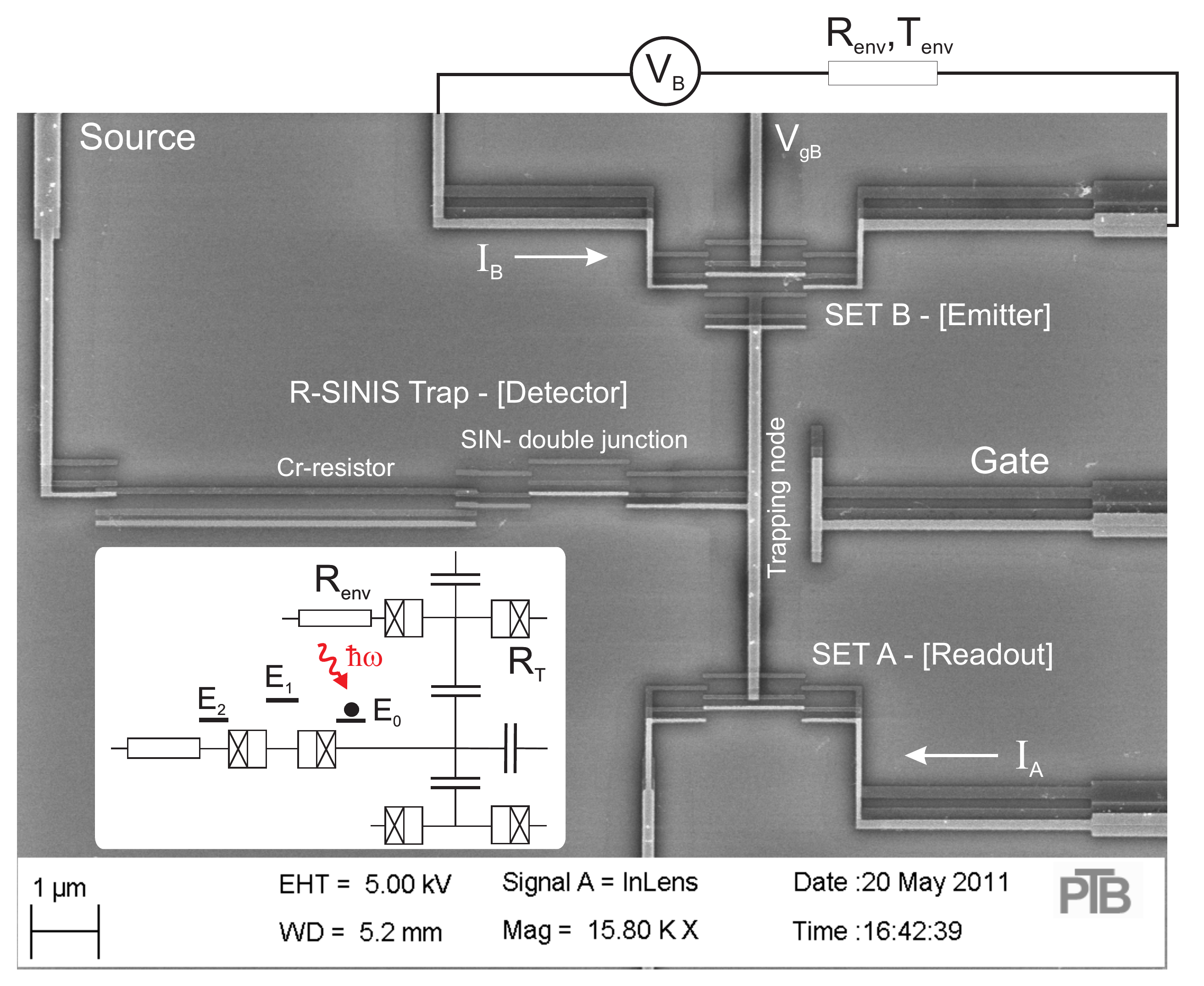}
\caption{
(Color online) Scanning electron micrograph of the sample, fabricated in a three-shadow deposition sequence, including Cr resistor
and S(Al) and N(AuPd) leads ordered from top to bottom.
Inset: Equivalent circuit projecting the SIN tunnel junctions (double boxes with the crossed item for S- and the open
item for N-leads), Cr resistor, an effective impedance of the environment of the emitter SET B,
$R_{\rm {env}}$, and symbolically: propagation of photons towards the trap (Detector).
The short horizontal steps depict the electrostatic barrier for the electron charge captured in the trapping node.
}
\label{Fig1}
\end{figure}

In this Letter, we suggest a mechanism of the SET backaction which is
directly related to the dissipative properties of its electromagnetic
environment and the photon exchange between interacting single-electron
devices. In particular, we demonstrate operation of an electron trap as a
detector of absorbed photons generated on chip by an adjacent SINIS SET. A
two-transistor arrangement was used for this purpose with the R-SINIS trap
marked as "Detector" in Fig.~\ref{Fig1}. The trap was, on the one hand,
optimally coupled to a readout SET A biased at a small probing current
$I_{\rm A}$ and, on the other hand, to an emitter SET B biased at a
current $I_{\rm B}\gg I_{\rm A}$ as used to stimulate the photon-activated
electron escape in the trap.

\begin{figure}[t]
\centering%
\includegraphics[width=1\columnwidth]{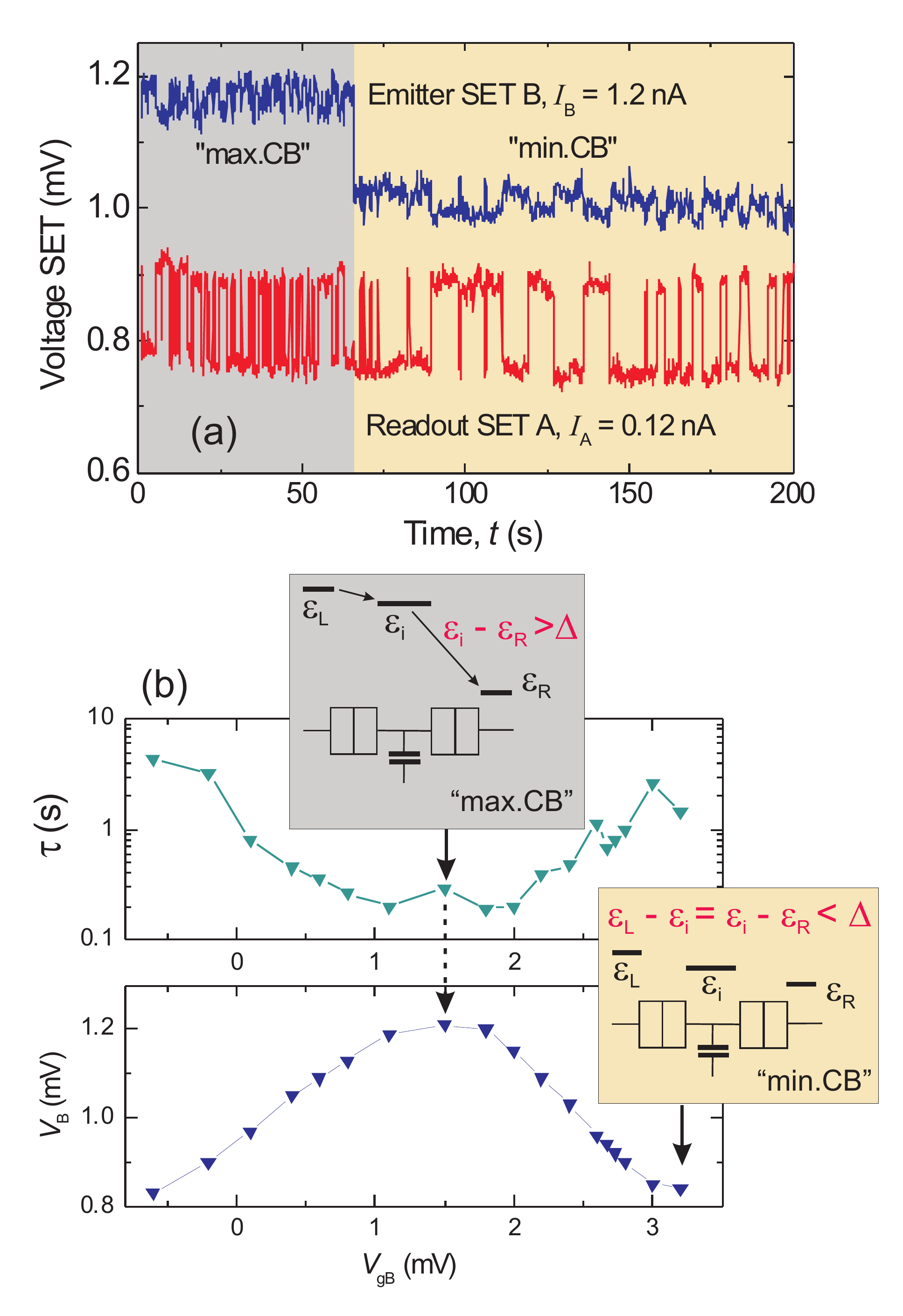}
\caption{
(Color online) Effect of the emitter SET B on the hold times of the trap. (a) Time traces taken simultaneously for SETs A and B.
The electrostatic barrier was symmetrized for achieving the similar average dwell times (=hold times) for the upper and lower charge states, see the trace for SET A.
At $t =$~65~s, by otherwise unchanged trapping and detection conditions (as proven by the signal from SET A),
the offset charge of SET B changed spontaneously giving rise to a lower value of $V_{\rm B}$, resulting in noticeably less frequent state switchings of the trap.
(b) Correlated behaviour of both voltage $V_{\rm B}$ and hold time $\tau$ in response to a modulation of the gate voltage $V_{\rm {gB}}$.
The bias currents were $I_{\rm A} = $~0.1~nA and $I_{\rm B} = $~1~nA.
Two energy pictographs, "min.CB" and "max.CB", correspond to the gate regimes of SET with a minimum and maximum Coulomb blockade, respectively.
In the state "max.CB" one of two tunneling steps is capable to release a photon with the energy up to $\varepsilon_{\rm i} - \varepsilon_{\rm R} > \Delta$,
which is high enough to excite a quasiparticle when absorbed in the trap.
}
\label{Fig2}
\end{figure}
An electron escape from (see inset in Fig.~\ref{Fig1}) or tunneling to the
trapping node involves the creation of two quasiparticles in sequence; the
first one with the energy $E_{\rm {qp}}$ sufficiently high to overcome the
electrostatic barrier formed by the SINIS double junction: $E_{\rm {qp}} >
E_1-E_0 + \Delta$, where $\Delta$ is the superconducting energy gap in the
S-leads. The background ("dark") rate of escape strongly depends on the
barrier height and on the noise level in the measuring setup. The barrier
height is tunable by the source and gate voltages, from zero to the upper
limit approaching the charging energy $E_{\rm C} \equiv e^2/2C_{\rm
{sum}}$ of the trap, where $C_{\rm {sum}}$ is a total capacitance of the
small N-island in the SINIS double junction. For the measuring setup and
the samples in this work (in particular, due to the relatively high
resistance value of Cr microstrip $R_{\rm {Cr}} \sim$ 600~k$\Omega$), the
hold times of a few hundreds of seconds were registered at the highest
barrier settings. Also, impractically high values of $I_{\rm B}$ were
necessary to stimulate the escape process, unless the electrostatic
barrier was tuned close to zero, $E_1 \approx E_0$, the quasiparticle
states involved close to the bottom of the excitation band  $E_{\rm {qp}}
\approx \Delta$, and the dark hold times were reduced by two orders of
magnitude down to the convenient experimental scale of a few seconds.

\begin{figure}[b]
\centering%
\includegraphics[width=1.05\columnwidth]{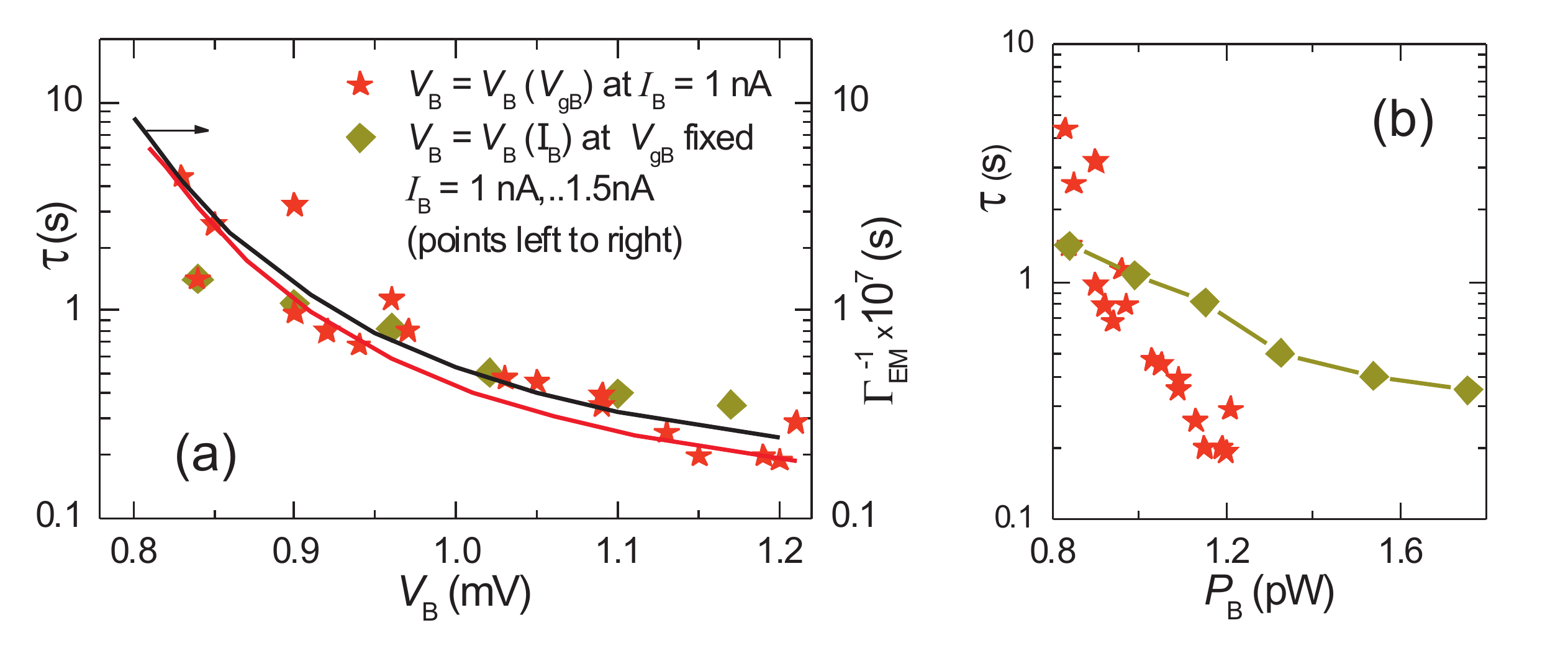}
\caption{
(Color online) (a) Two different data sets with the hold times $\tau$ plotted as a function of voltage $V_{\rm B}$.
(b) The same hold times plotted vs. total power $P_{\rm B}$ dissipated in the emitter SET B.
Two sets of measurements are shown: (1) $\tau(V_{\rm B})$ obtained at a fixed bias current $I_{\rm B}=$~1~nA by varying the gate voltage $V_{\rm {gB}}$ (stars);
(2) $\tau(V_{\rm B}, I_{\rm B})$ measured at fixed gate voltage (in the minimum of $V_{\rm B}$) by incrementing $I_{\rm B}=$~1,~1.1,...,~1.5~nA (diamonds).
The solid lines in (a) show the photon emission rates calculated with the following parameters: $T_{\rm e}=$~0.5~K, $E_{\rm SETB}=$~280~$\mu$eV,
$\Delta = $~250~$\mu$eV, $R_{\rm T}=$~150~k$\Omega$, where $E_{\rm SETB}$ is the charging energy of SET B defined in the same way as that of the trap. We model the environment of SET B
by $R_{\rm {env}}=$~1 and 1.25~k$\Omega$ and $T_{\rm {env}}=$~300 and 250~mK for the top and the bottom curves, respectively.
}
\label{Fig3}
\end{figure}

Under these conditions, we were able to observe a clear voltage dependence
of the hold time $\tau = \tau(V_{\rm B})$, where $V_{\rm B}$ is the
voltage across SET B. This effect is demonstrated in Fig.~\ref{Fig2}(a)
showing simultaneous time tracks for the voltage output of the readout SET
A and the emitter SET B. A random jump to an operating point with a lower
voltage $V_{\rm B} \sim$~1~mV resulted in noticeably less frequent state
switchings in the trap. This dependence was further verified by deliberate
gate modulation of the voltage $V_{\rm B} = V_{\rm B}(V_{\rm {gB}})$, at
fixed bias current $I_{\rm B}$, showing correlation with the hold times
$\tau(V_{\rm {gB}})$, see Fig.~\ref{Fig2}(b). In this measurement, the
direct electrostatic influence of the gate of SET B on the trap was
compensated by a cross-cancellation signal at the source terminal.
Remarkably, the maximum values of the hold times, $\tau \sim$~5~s in this
plot are close to the dark values with the emitter switched off, and the
minimum values are about an order of magnitude lower.

In order to draw conclusions about the direct heat transfer through the
substrate (cf., e.g., Ref.~\cite{Krupenin1999}), we plotted the same hold
time data as a function of either voltage $V_{\rm B}$, see
Fig.~\ref{Fig3}(a), or the dissipated power $P_{\rm B} \equiv V_{\rm B}
\times I_{\rm B}$ shown in Fig.~\ref{Fig3}(b). Two data sets measured: the
fixed-current data set and the incremental-current data do not collapse
into a single curve along the power axis in the plot (b), but they do
collapse along the $V_{\rm B}$-axis in the plot (a). Moreover, the slower
ramping within the incremental-current data set in Fig.~\ref{Fig3}(b)
would act counterintuitive, should we account for a more intensive heating
due to the increasing current $I_{\rm B}$. A comparison with
Fig.~\ref{Fig3}(a) allows us to conclude that rather the voltage $V_{\rm
B}$, but not the dissipated power $P_{\rm B}$ is a relevant parameter
describing the influence of the emitter SET B on the trap.

We interprete the dependence $\tau(V_{\rm B})$ following the argument of
the Environment-Assisted Tunneling (EAT) approach developed in
Ref.~\cite{Pekola2010}. We attribute the state switchings of the trap to
absorption of photons with energy $\hbar \omega \ge \Delta \sim
h\times$60~GHz first released during the process of electron tunneling in
SET B into the biasing leads and the other circuit components constituting
the electromagnetic environment (bath) of SET B \cite{IngNaz}. For the
purpose of modeling, we assumed an energy-independent trap switching
probability per incident photon which is obviously much lower than unity.

The photon emission rate was estimated using the standard master equation
treatment of SET B in a simplified form of a two-state approximation, see
the symbolic pictographs in Fig.~\ref{Fig2}(b): The basic transport
algorithm is considered with one tunneling event occurring in each of two
junctions in sequence. The emission rate $\Gamma_{\rm {EM}}$ is found as a
fraction $\gamma^{\rm {in,out}} \equiv \gamma(E^{\rm {in,out}})$ of the
corresponding tunneling rate $\Gamma^{\rm {in,out}} \equiv \Gamma(E^{\rm
{in,out}})$, where $E^{\rm {in,out}}$ is a free energy difference for the
incoming and the consequent outgoing tunneling event to/from the SET
island, respectively:
\begin{equation}
\Gamma_{\rm {EM}} = \left(\frac{\gamma^{in}}{\Gamma^{in}} + \frac{\gamma^{out}}{\Gamma^{out}} \right) \times \frac{I_{\rm B}}{e},
\label{relrate}
\end{equation}
where
\begin{eqnarray}
\Gamma(E)\left[ \gamma((E)\right] = \frac{1}{e^2R_T}
\int_{\Delta \left[2\Delta \right]}^\infty dE_{\rm n}
f_{\rm n}(E_{\rm n}-E,T_{\rm e})
\nonumber\\
\times
\int_{\Delta}^{\infty \left[ E_{\rm n}-\Delta \right]} dE_{\rm s}
n_{\rm s}(E_{\rm s})
P(E_{\rm n}-E_{\rm s}).
\label{goldenrule}
\end{eqnarray}
Here we use the standard Fermi factor in the N-island as $f_{\rm
n}(E,T_{\rm e}) = \left[1+exp(E/k_{\rm B}T_{\rm e})\right]^{-1}$ and
assume the quasiparticle band population to be negligible in the long
S-leads described by BCS density of states: $n_{\rm s}(E) =
\frac{E}{\sqrt{E^2 - \Delta^2}}$. The electron overheating in the
1.5~$\mu$m-long N-island was accounted for, using the temperature $T_{\rm
e}$ as a sensitive fitting parameter. The device parameters are taken
directly from the measurements. For the spectral function $P(E)$, we use
its gamma-function representation developed in
Ref.~\cite{IngoldGrabertEberhardt} for a small, semi-phenomenological
frequency-independent environmental impedance $R_{\rm {env}} \ll R_{\rm Q}
\equiv h/e^2 \approx \unit[25.8]{k\Omega}$.

Two solid lines in Fig.~\ref{Fig3}(a) show two possible fits, with
slightly different values of $R_{\rm {env}}$ and the effective temperature
$T_{\rm {env}}$, which appear realistic with respect to the sample design.
The calculated rates mimic the experimental dependency $\tau(V_{\rm B})$,
whereas the absolute value of the emission rate is obviously higher than
the experimental values of $\tau^{-1}$. This obvious effect arises due to
considerable thermal losses in the leads and a very weak photon coupling
to the trap, being, as a structure, much smaller than the photon
wavelength $\lambda \sim$~1~cm.

We also consider the RTN model of the backaction
\cite{Kemppinen2011,Saira2012}, viewing the SET island as a two-level
fluctuator (TLF) and valid even in the case $R_{\rm {env}} \to 0$. This
model however fails to explain the effect of the gate voltage shown in
Fig.~\ref{Fig2}, because, according to the TLF model (cf. Eqs.~(12,13) in
Ref.~\cite{Machlup1954}), the high-frequency tail of the RTN spectrum at
$\hbar \omega \ge \Delta$, responsible for the trap excitations, should
not be sensitive to the duty cycle of TLF distinguishing the different
gate regimes. We note that a vanishing contribution of the RTN model might
be a result of a rapid decay, $P_{\rm t}(E) \propto E^{-4}$ for $E <0$
\cite{Martinis1993}, of the noise-related spectral $P_{\rm t}$-function of
the trap.

In conclusion, our analysis of the tunneling process in SINIS SET
unambiguously shows proportionality between the photon emission rate and
the state switching frequency of the R-SINIS trap thus operated as a
microwave photon detector. Non-zero environmental impedance of the
photon-emitting SET plays an important role in the interaction process
with the trap. More detailed study is necessary on the on chip propagation
of the photons within a dedicated circuitry as well as on the detailed
spectrometric function of the trap. Of metrological interest could be a
mutual accuracy impact due to the photon exchange in an array of hybrid
turnstiles operating in parallel \cite{10turnstiles}.

We acknowledge useful discussions with J.~P. Pekola and A. Kemppinen and
experimental support from T. Weimann and V. Rogala. The research conducted
within EU project SCOPE has received funding from the European Community's
Seventh Framework Programme under Grant Agreement No. 218783.


\begin{thebibliography}{13}

\bibitem{Pekola2008} J.~P. Pekola, J.~J. Vartiainen, M. M\"{o}tt\"{o}nen,
    O.-P. Saira, M. Meschke, and D.~V. Averin, Nature Phys. {\bf 4},  120  (2008).

\bibitem{Camarota2012} B. Camarota, H. Scherer, M.~W. Keller, S.~V.
    Lotkhov, G.-D. Willenberg and F.~J. Ahlers, Metrologia {\bf 49}, 8-14 (2012).

\bibitem{Saira2012} O.-P. Saira, A. Kemppinen, V.~F. Maisi, and J.~P.
    Pekola, Phys. Rev. B {\bf 85}, 012504 (2012).

\bibitem{Kemppinen2011} A. Kemppinen, S.~V. Lotkhov, O.-P. Saira, A.~B.
    Zorin, J.~P. Pekola, and A.~J. Manninen, Appl. Phys. Lett. {\bf 99},
    142106 (2011).

\bibitem{Lotkhov-LT26} S.~V. Lotkhov and A.~B. Zorin, arXiv:1106.6005
    (2011).

\bibitem{Lotkhov2009} S.~V. Lotkhov, A. Kemppinen, S. Kafanov, J.~P.
    Pekola, and A.~B. Zorin, Appl. Phys. Lett. {\bf 95},  112507  (2009).

\bibitem{Lotkhov2011} S.~V. Lotkhov, O.-P. Saira, J.~P. Pekola, and A.~B.
    Zorin, New J. Phys. {\bf 13}, 013040 (2011).

\bibitem{Korotkov1994} A.~N.~Korotkov, Phys. Rev. B {\bf 49}, 10381
(1994).

\bibitem{Krupenin1999} V.~A. Krupenin, S.~V. Lotkhov, H.~Scherer, Th.~
    Weimann, A.~B. Zorin, F.-J. Ahlers, J. Niemeyer, and H. Wolf, Phys. Rev. B {\bf 59}, 10778 (1999).

\bibitem{10turnstiles} V.~F.~Maisi, Yu.~A.~Pashkin, S.~Kafanov,
    J.~S.~Tsai, and J.~P.~Pekola, New J. Phys. {\bf 11}, 113057 (2009).

\bibitem{Pekola2010} J.~P. Pekola, V.~F. Maisi, S. Kafanov, N. Chekurov,
    A. Kemppinen, Y.~A. Pashkin, O.-P. Saira, M. M\"ott\"onen, and J.~S. Tsai, Phys. Rev. Lett. {\bf 105},  026803  (2010).

\bibitem{IngNaz} G.~L.~Ingold and Yu.~V.~Nazarov, in \it {Single Charge
    Tunneling,} \rm edited by H.~Grabert and M.~H.~Devoret (Plenum, New
    York, 1992), Chap. 2.

\bibitem{IngoldGrabertEberhardt} G.~L.~Ingold, H. Grabert, and U. Ebert,
    Phys. Rev. B {\bf 50},  395  (1994).

\bibitem{Machlup1954} S.~Machlup, J. Appl. Phys. {\bf 25}, 341 (1954).

\bibitem{Martinis1993} J.~M.~Martinis and M.~Nahum, Phys. Rev. B {\bf 48},
    18316 (1993).

\end{thebibliography}
\end{document}